\documentclass[Numbered,pdflatex,sn-nature]{sn-jnl}

\usepackage{graphicx}%
\usepackage{caption}

\usepackage{multirow}%
\usepackage{amsmath,amssymb,amsfonts}%
\usepackage{amsthm}%
\usepackage{mathrsfs}%
\usepackage[title]{appendix}%
\usepackage{xcolor}%
\usepackage{textcomp}%
\usepackage{manyfoot}%
\usepackage{booktabs}%
\usepackage{algorithm}%
\usepackage{algorithmicx}%
\usepackage{algpseudocode}%
\usepackage{listings}%
\usepackage{threeparttable} 

\def\farcm{%
 \mbox{.\kern -0.7ex\raisebox{.9ex}{\scriptsize$\prime$}}%
}%
\def\farcs{%
 \mbox{%
 \kern 0.13ex.%
 \kern -0.95ex\raisebox{.9ex}{\scriptsize$\prime\prime$}%
 \kern -0.1ex%
 }%
}%

\theoremstyle{thmstyleone}%
\theoremstyle{thmstyletwo}%

\theoremstyle{thmstylethree}%

\begin{document}

\title[FL$\alpha$RE-z7: the First Ly$\alpha$ transmission in the Reionization Epoch at $z\sim7$]{A 210 Megaparsec Ionized Structure at $z=6.8$ Revealed by an Astrophysics Foundation Model}


\author*[1]{\fnm{Minghao} \sur{Yue}}\email{yuemh@arizona.edu}
\equalcont{These authors contributed equally to this work.}

\author*[1]{\fnm{Yongda} \sur{Zhu}}\email{yongdaz@arizona.edu}
\equalcont{These authors contributed equally to this work.}

\author[1]{\fnm{Jiani} \sur{Ding}}\email{jianiding@arizona.edu}
\equalcont{These authors contributed equally to this work.}

\author[1]{\fnm{Xiaohui} \sur{Fan}}\email{xfan@arizona.edu}

\author[2]{\fnm{Yufeng} \sur{Luo}}\email{yluo4@uwyo.edu}


\affil*[1]{\orgdiv{Steward Observatory}, \orgname{University of Arizona}, \orgaddress{\street{933 N. Cherry Ave.}, \city{Tucson}, \postcode{85721}, \state{Arizona}, \country{USA}}}

\affil[2]{\orgdiv{Department of Physics and Astronomy}, \orgname{University of Wyoming}, \orgaddress{\street{1000 E. University Ave.}, \city{Laramie}, \postcode{82071}, \state{Wyoming}, \country{USA}}}

\abstract{
Reionization is the last major transition in the Universe that transforms the intergalactic medium (IGM) from neutral to highly ionized, yet its spatial structure remains poorly constrained. Using an astrophysics foundation model trained for James Webb Space Telescope datasets, we identify a luminous galaxy at redshift $z=7.567$ showing significant Ly$\alpha$ transmission at $z=6.80\pm0.07$. This transmission, named ``the First Ly$\alpha$ transmission in the Reionization Epoch at $z\sim7$" (FL$\alpha$RE-z7), indicates a cosmological-scale ionized structure extending over $210.5^{+55.7}_{-53.5}$ comoving Megaparsecs at cosmic dawn. The hydrogen neutral fraction of the ionized structure is $x_{\rm HI}=(2.07^{+0.29}_{-0.30})\times10^{-6}$, about $10^5$ times lower than the cosmic mean at similar redshifts $(x_{\rm HI}\approx0.4)$. Cosmological simulations suggest a probability of $\approx10^{-5}$ for a random sightline to reproduce the transmission. We find no significant galaxy overdensity associated with the transmission region, in tension with canonical inside-out reionization models. FL$\alpha$RE-z7 severely challenges existing reionization models in the standard $\Lambda$CDM Universe, indicating extreme spatial fluctuation in ionizing radiation or IGM temperature. Our result marks the first example of a foundation model discovering a previously unpredicted astrophysical phenomenon, demonstrating the power of deep learning methods for mining large astronomical datasets and finding unexpected new astrophysics.}

\keywords{High-redshift Galaxies, Reionization, Intergalactic Medium, Deep learning}



\maketitle

\section{Main}

Deep learning and Artificial Intelligence (AI) are revolutionizing the way scientific discoveries are made across a wide variety of disciplines 
\cite{Jumper2021,Merchant2023,Lam2023,Degrave2022}. Among these advances, foundation models -- a family of self-supervised learning networks that learn the structure of large datasets without class labels -- are especially suitable for identifying previously unknown objects and phenomena (e.g., \cite{Bommsani2021,Hayat2021}). In astronomy and astrophysics, several foundation models have been recently developed (e.g., \cite{Parker2024AstroCLIP,Parker2025,Euclidfm2026}), yet these studies focused on surveying known source populations and refining property measurements of known objects. Whether deep learning and foundation models can reveal unknown or unexpected astrophysical phenomena remains poorly explored. Answering this question is critical for advancing our understanding of the Universe, as ongoing and upcoming sky surveys are expected to deliver large numbers of ``unknown unknowns'', which might encode new astrophysics and cannot be identified using pre-defined source selection algorithms.

In this work, we report the first foundation-model-driven discovery of an unpredicted astrophysical phenomenon: an ionized cosmological-scale structure in the epoch of reionization (EoR) when the Universe is substantially neutral. This ionized superstructure, which we name ``the First Ly{$\bf{\alpha}$} transmission in the Reionization Epoch at $z\sim7$'' (FL$\alpha$RE-z7), is not predicted by cosmological simulations and challenges our understanding of the early Universe and reionization. 
Reionization is the last major phase transition of the Universe, where the hydrogen in the intergalactic medium (IGM) turns from neutral to ionized from redshift $z\gtrsim10$ to $z\approx6$ \cite{fan06,Yang2020,bosman_hydrogen_2022}. Reionization affects about 90\% of the baryonic matter in the Universe and has profound influences on galaxy evolution through cosmic time, yet its spatial structure remains uncertain. Previous models and observations of reionization suggested an IGM neutral fraction of $x_{\rm HI}\approx0.4$ around redshift $z\approx7$ \cite{jin23darkpixel,durovcikova24}. In contrast, FL$\alpha$RE-z7 exhibits a neutral fraction of $x_{\rm HI}\approx10^{-6}$ at about 200 comoving megaparsecs (Mpc) scales, which cannot be explained by canonical models of reionization in the standard $\Lambda$CDM Universe. FL$\alpha$RE-z7 implies a highly patchy spatial structure for reionization, and its discovery demonstrates the ability of foundation models to discover unknown and unpredicted astrophysical phenomena.

We would like to note that JADES-GS-217704 is independently and simultaneously reported by \cite{helton2026LAF}, where the object is identified by human inspection and named JADES-GS-z7-LAF by the authors.

\subsubsection*{A Foundation-Model-Based Framework Leads to the Discovery of FL$\alpha$RE-z7}

We use FM-JADES-v1 \cite{ding26learningjwst1}, a foundation model for James Webb Space Telescope (JWST) extragalactic studies, to develop discovery frameworks for new astrophysical objects and phenomena. Here we briefly summarize the information about FM-JADES-v1 and refer readers to \cite{ding26learningjwst1} for detailed information. FM-JADES-v1  is trained on the JWST Advanced Deep Extragalactic Survey (JADES) data release 5 (DR5) imaging and photometry dataset \cite{Carreira2026,Eisenstein2026,Johnson2026,Robertson2026}. For each object in JADES, the image patches and catalog columns are tokenized and enter a Transformer block \cite{Vaswai2017} that learns the relationships between different tokens. No class labels are assigned to objects during training. Each object is assigned a CLS token, which receives information from image and catalog tokens via the Transformer block and thus represents the object's features. We regard the CLS tokens of objects as their embedded vectors, which are used to investigate the distribution of objects in the representation space.

\begin{figure} 
	\centering
	\includegraphics[width=1\textwidth,trim={24pt 0 12pt 0}]{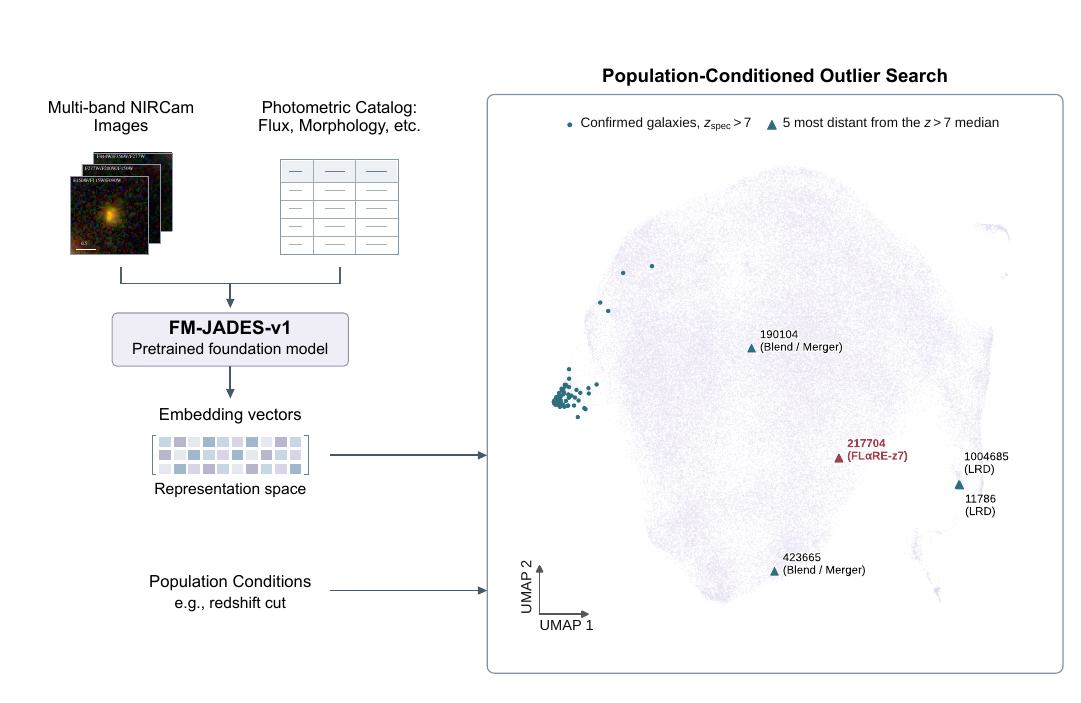}
	\caption{\textbf{Foundation-model-driven framework leads to the discovery of a high-redshift galaxy with extraordinary properties.} We feed the multi-band NIRCam images and photometric catalogs of 122,336 objects from the JADES DR5 dataset to the pre-trained astrophysics foundation model, FM-JADES-v1 \cite{ding26learningjwst1}. For each object, FM-JADES-v1 encodes the observational data into an embedding vector. We then project the embedding vectors onto a two-dimensional plane using UMAP \cite{umap}, and identify spectroscopically confirmed $z>7$ galaxies among this sample. We then perform the Population-Conditioned Outlier Search by identifying outliers that are far from the center of the $z>7$ galaxy distribution. These outliers are $z>7$ galaxies that exhibit anomalous properties. The main target of this study, JADES-GS-217704 (red triangle), is among the top five outliers in the high-redshift galaxy sample. The small magenta dots mark the 122,336 JADES objects we investigate, the blue dots are $z>7$ galaxies, and the triangles are the top five outliers.}
	\label{fig:method} 
\end{figure}

In this work, we introduce the {\em population-conditioned outlier search} framework for accelerated discovery of objects with unusual or unexpected properties. Specifically, we start from a pre-defined population with certain astrophysical properties, obtain the embedding vector for these objects, and search for outliers in the embedded space to identify objects with extraordinary features (Figure \ref{fig:method}, left).
As an application of this framework, this paper searches for high-redshift galaxies $(z>7)$ with unusual properties in the JADES survey.
To this end, 
we use the frozen FM-JADES-v1 model to produce the embedded vector for 122,336 objects in JADES, which are uniformly selected to have coverage of at least seven wide JWST NIRCam filters (see \cite{ding26learningjwst1,methods} for details). The right panel of Figure \ref{fig:method} shows the distribution of these objects in the projected UMAP space \cite{umap}. We set the population condition to be ``spectroscopically confirmed galaxies at $z>7$'' and investigate their distribution in the representation space  (Methods). Most of the $z>7$ galaxies are concentrated on the left side of the UMAP, forming a high-redshift island in the representation space (see \cite{ding26learningjwst1} for details). Meanwhile, a few $z>7$ galaxies are scattered across other regions on the UMAP. We compute the distance between each object and the median of all $z>7$ galaxies, and rank the objects by their distance to the sample median. Top-ranked objects are expected to have unusual astrophysical features compared to the main high-redshift galaxy population. Following the ranked list, for each object, we feed the observations to Large Language Models (LLMs) including Claude \cite{anthropic_claude_2026} and ChatGPT\cite{openai_chatgpt_2026} to generate hypotheses for the origin of each anomaly. At the final stage of the framework, the authors investigate the top outliers and hypotheses to identify potentially new astrophysical phenomena.

In the final ranked object list, the top two are Little Red Dots (LRDs), a population of active galactic nuclei recently discovered by JWST \cite{Matthee2024}. LRDs have distinct morphology and spectral energy distribution compared to other galaxies, which explains why these two objects are outliers. The two LRDs have been reported by previous studies (e.g., \cite{Rinaldi2025}). The third one, JADES-GS-217704 $(z=7.567)$, exhibits an unusual feature: there is significant flux bluer than its Ly$\alpha$ wavelength, where we expect non-detection due to neutral IGM attenuation. 
In the rest of the study, we will focus on the properties of JADES-GS-217704. More discussion of other outliers can be found in Methods.

\begin{figure} 
	\centering
	\includegraphics[width=1.0\textwidth]{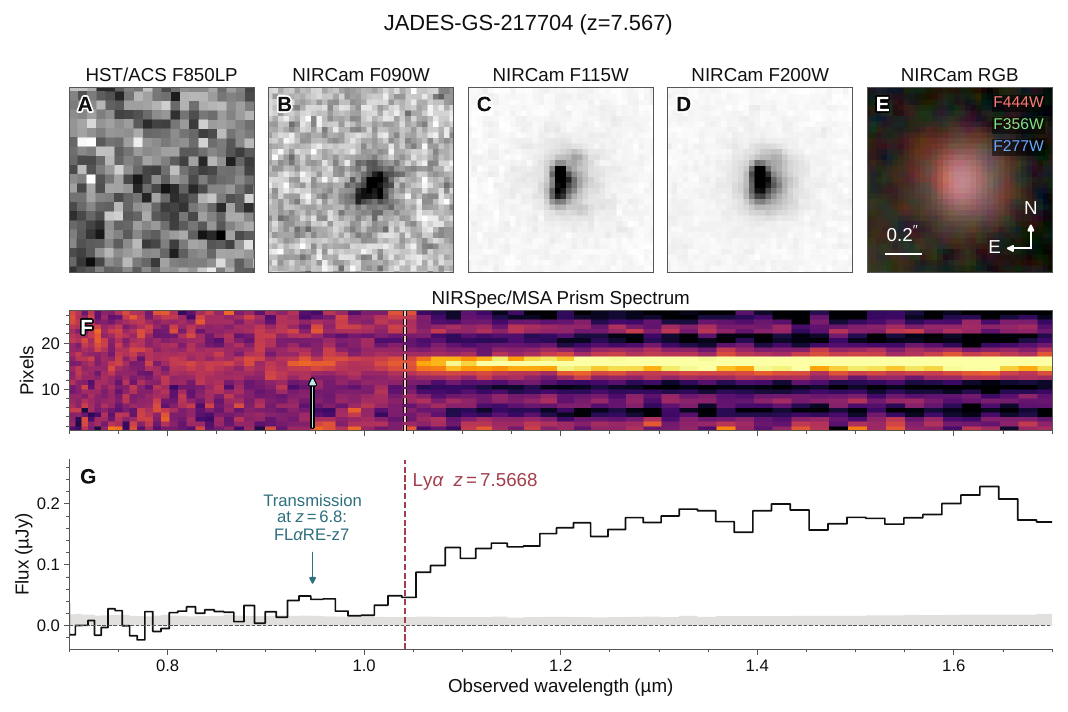}
	\caption{\textbf{HST and JWST observations for JADES-GS-217704.} 
        (\textbf{A}) The HST F850LP image. This object is undetected in deep optical imaging by HST. (\textbf{B}) NIRcam F090W imaging presents a $13\sigma$ detection, indicating integrated flux at wavelengths blueward of Ly$\alpha$ emission. (\textbf{C-D}) NIRcam F115W and F200W images. (\textbf{E}) NIRCam F277W, F356W, and F444W images, shown in the RGB format. (\textbf{F} and \textbf{G}) JWST NIRSpec/MSA Prism 2D spectrum and 1D spectrum. The red dashed line denotes the position of the Ly$\alpha$ break, and the teal arrow marks the position of the transmission feature (FL$\alpha$RE-z7). The gray-shaded region in the 1D spectrum indicates the $1\sigma$ error. The 1D spectrum indicates a $5.66\sigma$ detection between $0.9\mu{\rm m}$ and $1\mu{\rm m}$.}
	\label{fig:spectra} 
\end{figure}

Figure \ref{fig:spectra} shows the Hubble Space Telescope (HST) image, JWST NIRCam images, and NIRSpec/MSA Prism spectra of JADES-GS-217704 (see Methods for details about the observations). JADES-GS-217704 is a luminous galaxy at $z_{\rm spec}=7.567$ with $m_{\rm F115W}=26.17 \pm 0.01$. The redshift is securely determined by H$\beta$ and [O\,{\sc III}] emission lines (Methods). Its Ly$\alpha$ wavelength is $\lambda_{{\rm Ly}\alpha}=1.042\mu$m, redder than the coverage of the F090W filter (0.795 -- 1.005 $\mu$m)\footnote{https://jwst-docs.stsci.edu/jwst-near-infrared-camera/nircam-instrumentation/nircam-filters\#gsc.tab=0}. Meanwhile, JADES-GS-217704 has a significant F090W detection of $m_{\rm F090W}=28.20 \pm 0.08$, indicating positive flux bluer than its Ly$\alpha$ wavelength. The NIRSpec/MSA Prism spectrum also exhibits significant positive flux at $\lambda_{\rm obs}\approx0.95\mu$m; when integrated between $0.9{\rm \mu m}<\lambda_{\rm obs}<1.0{\rm \mu m}$, this feature has a significance of $5.66\sigma$. The significant flux bluer than $\lambda_{{\rm Ly}\alpha}$ contradicts the standard picture of cosmic reionization, which suggests that the IGM at $z\approx7$ has high neutral fractions of $x_{\rm HI}\approx0.4$ and should absorb all rest-frame ultraviolet (UV) photons bluer than $\lambda_{{\rm Ly}\alpha}$ for $z\gtrsim7$ galaxies. To produce this feature, there must be an ionized structure on the sightline of JADES-GS-217704 at $z\approx6.8$ with a very low hydrogen neutral fraction $(x_{\rm HI}\lesssim10^{-6})$. We refer to this ionized structure as ``the First Ly{$\bf{\alpha}$} transmission in the Reionization Epoch at $z\sim7$'' (FL$\alpha$RE-z7). We note that the flux cannot come from a foreground interloper, given its HST photometry and the spectral shape at $\lambda_{\rm obs}<1\mu$m (see Methods for details).

The discovery of FL$\alpha$RE-z7 challenges canonical models of the early Universe. Below we determine the properties of FL$\alpha$RE-z7 and discuss its implications for how reionization happens.

\subsubsection*{A Cosmological-Scale Ionized Superstructure in the Neutral Early Universe}

\begin{figure} 
	\centering
	\includegraphics[width=0.7\textwidth]{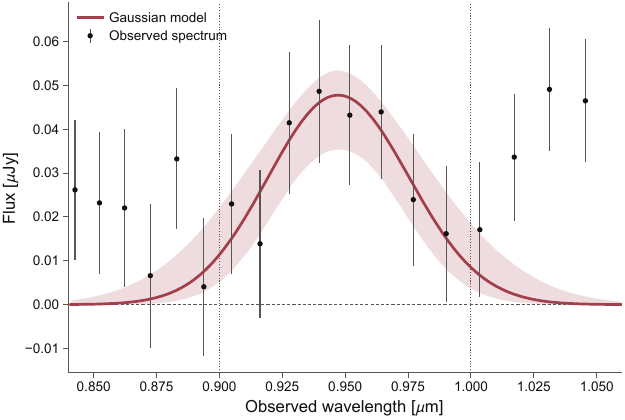}
	\caption{\textbf{Determining the redshift and the width of the transmission signal.}
		We fit a Gaussian profile for the transmission signal, adopting the LSF from \cite{adg24}. The dotted lines mark the fitting window, and the shaded region marks the 1$\sigma$ posterior of the Gaussian model. The model suggests a central wavelength of $\lambda_{\rm c}=0.948\pm0.008\mu{\rm m}$, corresponding to a redshift of $z_{\rm c}=6.80\pm 0.07$. The FWHM of the transmission is $0.07\pm0.02\mu$m, corresponding to $210.5^{+55.7}_{-53.5}$ cMpc.}
	\label{fig:size}
\end{figure}

To determine the redshift and size of the ionized structure, we fit the transmission signal between $0.9\mu$m and $1\mu$m as a Gaussian profile. We adopt the NIRSpec/MSA Prism Line Spread Function (LSF) from \cite{adg24}. Figure \ref{fig:size} shows the best-fit model, which indicates a central wavelength of $\lambda_{\rm c}=0.948\pm0.008\mu{\rm m}$ and a FWHM of $0.07\pm0.02\ \mu\mathrm{m}$. These results translate to a redshift of $z=6.80\pm 0.07$ and a size of $210.5^{+55.7}_{-53.5}$ comoving Megaparsec (cMpc) for the ionized structure (Methods). 

\begin{figure} 
	\centering
	\includegraphics[width=0.8\textwidth]{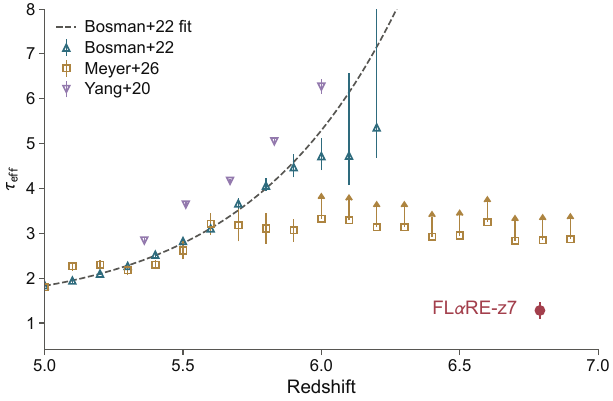}
	\caption{\textbf{The IGM Ly$\alpha$ optical depth as a function of redshift.} We include IGM optical depth constraints from high-redshift quasar and galaxy observations \cite{Yang2020,bosman_hydrogen_2022,meyer_probing_2025}.
		The IGM at $z\approx5.5$ has optical depth $\tau_{\rm eff}\approx2$, which quickly increase to $\tau_{\rm eff}>4$ at $z\gtrsim6.5$. The optical depth FL$\alpha$RE-z7 $(\tau_{\rm eff}=1.28\pm0.18)$ is similar to the cosmic mean value at $z\approx5$, indicating near-complete reionization progress for FL$\alpha$RE-z7 despite its high redshift.}
	\label{fig:tau}
\end{figure}

We also evaluate the effective Ly$\alpha$ optical depth for two wavelength windows centered at $\lambda_{\rm c}$, with widths of FWHM and 50 cMpc$/h$, respectively. We use the 50 cMpc$/h$ window following previous studies of the reionization-era IGMs \cite{eilers_opacity_2018,Yang2020,jin_probing_2026}. We reconstruct the intrinsic continuum of the galaxy by fitting a spectral energy distribution (SED) using \texttt{Prospector} \cite{prospector} 
(see details in Methods). 
Our analysis yields an effective optical depth of $\tau_{\rm eff}=1.28\pm0.18$ for the FWHM window and $1.08\pm0.24$ for the 50 cMpc$/h$ window. 
The corresponding hydrogen neutral fraction of $x_{\rm HI}=(2.07^{+0.29}_{-0.30})\times10^{-6}$ and $(1.74^{+0.38}_{-0.39})\times10^{-6}$, respectively (Methods). 
FL$\alpha$RE-z7 has an optical depth comparable $z\approx5$ IGM, which is much lower than the $\tau_{\rm eff}$ at $z>6.5$ measured by previous studies (Figure \ref{fig:tau}).

At $z\approx6.8$ (about 800 Myrs after the Big Bang), the Universe is highly neutral, with an IGM hydrogen neutral fraction of $x_{\rm HI}\approx0.4$ \cite{Yang2020,jin_probing_2026,zhu_galaxy_2026}. Such a high neutral fraction produces complete Gunn-Peterson absorption troughs \cite{gunnpeterson} for $z>7$ galaxies, leaving no flux bluer than their Ly$\alpha$ wavelengths. Although ionized bubbles at $z > 7$ have been discovered by identifying Ly$\alpha$ emission lines \cite{willott25,witstok25LAE}, these ionized bubbles are small (with sizes of $\lesssim10$ cMpc) and are localized around luminous galaxies or active galactic nuclei (AGN).
In contrast, FL$\alpha$RE-z7 has a size of $\gtrsim200$ cMpc, much larger than ionized bubbles produced by galaxies and even by the most luminous quasars \cite{Eilers2017}. We also check objects in the JADES DR5 catalog around the FL$\alpha$RE-z7 sightline, and found no luminous objects ($m_{\rm F115W}<24$) within 3 arcminutes with photometric redshift $|z_{\rm phot}-z_{\rm c}|<0.2$. We therefore conclude that FL$\alpha$RE-z7 is an ionized superstructure at cosmological scales, rather than an ionized bubble produced by a single luminous object.

\subsubsection*{FL$\alpha$RE-z7 Challenges Existing Models for the Early Universe}

The existence of a highly-ionized superstructure like FL$\alpha$RE-z7 challenges current models of the early Universe, in particular for its reionization history. We demonstrate this point by comparing the optical depth of FL$\alpha$RE-z7 with cosmological simulations in \cite{keating20reionization}.
To consider the impact of different astrophysical parameters on reionization,
\cite{keating20reionization} considered three models, namely the high cosmic wave background optical depth $(\tau_{\rm CMB})$ model, the low $\tau_{\rm CMB}$ model, and the hot low $\tau_{\rm CMB}$ model. A high $\tau_{\rm CMB}$ indicates earlier reionization, and the hot model means higher energy for the ionizing photons. For each model, \cite{keating20reionization} provided the Ly$\alpha$ optical depths for 20,000 skewers, and we evaluate the $\tau_{\rm eff}$ for these skewers using the same wavelength window as for FL$\alpha$RE-z7.

We evaluate the probability distribution of $\tau_{\rm eff}$ for the simulated skewers. We use the continuum model from SED fitting as the background source spectrum at $z=7.567$ and apply the IGM attenuation of each skewer on the model continuum. We then compute the transmission flux for the 50 cMpc$/h$ centered at FL$\alpha$RE-z7's redshift (Methods).
All three models fail to produce sightlines with $\tau_{\rm eff}<4$. 
After including measurement errors, we estimate that the probability for a random sightline to have a measured flux higher than FL$\alpha$RE-z7 is $\approx10^{-5}$ for all three models. Since FL$\alpha$RE-z7 is discovered among 100 spectroscopically confirmed $z>7$ galaxies, the probability of finding such a transparent sightline in our entire experiment is about $10^{-3}$. 

This result indicates a tension between the discovery of FL$\alpha$RE-z7 and the existing cosmological simulations for reionization. This tension cannot be explained by an early reionization model, as the high-$\tau_{\rm CMB}$ model fails to produce such a sightline. The most plausible solution is a highly patchy reionization model, where the ionizing radiation or IGM temperature has large fluctuations at $\gtrsim50$ cMpc scales. This picture is consistent with the $\gtrsim100$ cMpc neutral IGM structures at the end of reionization ($z\approx5.5$), when the Universe has been largely ionized \cite{Davies2018LongGap,zhu2022darkgap}.

\subsubsection*{No Evidence for Significant Galaxy Overdensities around FL$\alpha$RE-z7}

In the canonical ``inside-out" picture of reionization, large ionizing structures at $z\gtrsim6$ should reside in galaxy overdensities or protoclusters \cite{neyer_thesan_2024,Lu2024Bubble}, as copious amounts of ionizing photons from galaxies are required to produce large ionized bubbles. Nevertheless, this canonical picture has been challenged by recent cosmological simulations \cite{garaldi2025_outsidein}, which suggested that the IGM density in galaxy overdensities is too high to get ionized. Instead, the simulations support the ``outside-in" reionization, where ionized structures emerge from underdense regions in the Universe. 

\begin{figure}
	\centering
	\includegraphics[width=1\textwidth]{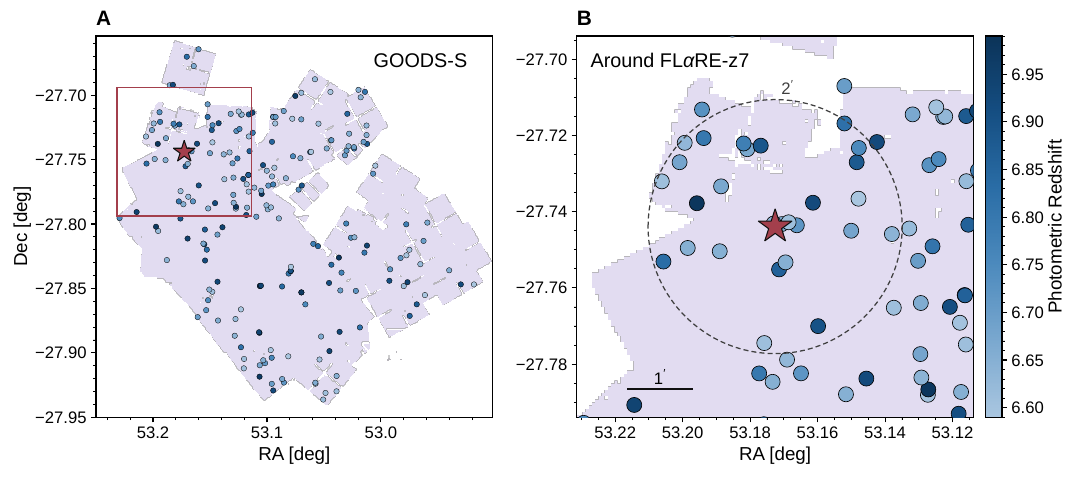}
	\caption{\textbf{The environment of FL$\alpha$RE-z7.} {\bf (A)} the GOODS-S field; {\bf (B)} the region around FL$\alpha$RE-z7, corresponding to the red box in the left panel. The background shaded area shows the footprint that reaches a given set of depth criteria (defined in Methods); the dots are objects brighter than the depth criteria in the JADES DR5 catalog, colored by their photometric redshifts.
		We identify objects in this footprint with photometric redshifts $|z_{\rm phot}-6.8|<0.2$. The entire field has an object surface number density of $\Sigma=1.44\pm0.10 {\rm ~arcmin^{-2}}$, and the area within $2'$ around FL$\alpha$RE-z7 has $\Sigma=2.28\pm0.44{\rm ~arcmin^{-2}}$. In other words, there is no evidence for an overdensity around FL$\alpha$RE-z7.}
	\label{fig:environ} 
\end{figure}

FL$\alpha$RE-z7 provides a direct test of the two pictures about reionization. Figure \ref{fig:environ} shows the distribution of objects with $|z_{\rm phot}-z_{\rm c}|<0.2$ in the JADES GOODS-S field. We only count areas with at least seven NIRCam wide filters coverage and apply magnitude cuts to ensure a uniform coverage across the field (see Methods for details). The average object surface number density is $1.44\pm0.10 {\rm ~arcmin^{-2}}$ across the entire GOODS-S field, and is $2.28\pm0.44{\rm ~arcmin^{-2}}$ within $r<2$  arcmin around FL$\alpha$RE-z7's sightline. 
In other words, there is no evidence for a strong overdensity around FL$\alpha$RE-z7.
This result indicates that ionized superstructures in the EoR do not appear to coincide with obvious concentrations of ionizing sources.
Similar behavior is seen near the end of reionization, where some of the clearest IGM Ly$\alpha$ transmission occurs in average-density or underdense regions \cite{christenson_relationship_2023,zhu_galaxy_2026}. Recent JWST observations likewise found higher galaxy Ly$\alpha$ visibility in lower-density environments over $4.8<z<11$ \cite{zhu_low_2026}.

The origin of such a large ionized structure without a strong galaxy overdensity remains unclear.  
Very metal-poor stellar populations can produce harder and more efficient ionizing spectra, and unusually high escape fractions or faint or obscured AGN could further increase the ionizing emissivity \cite{Schaerer2003}. 
Anisotropic escape provides another possibility: ionizing photons need not trace the projected galaxy distribution, and both observations and simulations suggest that directional escape can substantially alter the geometry of reionization \cite{gangolli_correlation_2025,zhu_galaxy_2026,Schwandt2026}. With a size of $\sim200$~cMpc, FL$\alpha$RE-z7 may consist of multiple overlapping ionized regions rather than a single dominant structure. Deeper imaging and spectroscopy of galaxies throughout and around the transmission region will shed light on the origin of this structure.


\subsection*{Summary}


We report the discovery of FL$\alpha$RE-z7, an ionized superstructure at $z=6.8$ with a line-of-sight size of $\sim200$ cMpc. FL$\alpha$RE-z7 is located on the sightline of a luminous $z=7.567$ galaxy, JADES-GS-217704, which shows significant positive flux bluer than its Ly$\alpha$ break in both NIRCam images and NIRSpec/MSA Prism spectrum. Using the transmission flux, we estimate a Ly$\alpha$ effective optical depth of $\tau_{\rm eff}=1.28\pm0.18$, corresponding to a neutral fraction of $x_{\rm HI}=(2.07^{+0.29}_{-0.30})\times10^{-6}$. FL$\alpha$RE-z7 provides the first observational evidence of near-complete reionization in such early cosmic times, when the Universe is overall substantially neutral. The environment of FL$\alpha$RE-z7 suggests that reionization can happen early even without strong galaxy overdensities. FL$\alpha$RE-z7 is not reproduced by cosmological simulations, challenging the canonical model of reionization in a $\Lambda$CDM Universe. Our interpretation for this system is largely consistent with that in \cite{helton2026LAF}.

The discovery of FL$\alpha$RE-z7 is driven by the foundation model FM-JADES-v1, which identifies the object as a $z>7$ galaxy with unusual observational features. We exploit the population-conditioned outlier search framework in our study, a novel approach of using foundation models for new phenomenon discovery. 
Notably, FM-JADES-v1 was not explicitly trained for IGM or reionization studies. This result emphasizes the potential of foundation models as general discovery machines for unknown or unexpected astrophysical phenomena.
As ongoing wide-field sky surveys (like the Euclid survey, the Roman Space Telescope Surveys, and the Vera C. Rubin Legacy Survey for Space and Time) expand object samples by orders of magnitude, foundation models offer a way to identify rare and unexpected physical signatures across these much larger datasets, opening a new path for discovery in all fields of astrophysics.


\section{Method}\label{sec:Method}



\subsubsection*{Cosmology and Magnitude System}

We adopt a Flat $\Lambda$CDM cosmology with $H_0=70{\rm~km~s^{-1} Mpc^{-1}}$ and $\Omega_{\rm M}=0.3$. All magnitudes in this work are AB magnitudes \cite{ABmag}.

\subsubsection*{HST and JWST Observations}

The images and photometric information used in this work are from JADES DR5. JADES DR5 provides NIRCam mosaic images and photometric catalogs in GOODS-S and GOODS-N fields \cite{Johnson2026,Robertson2026,Eisenstein2026}. The total coverage is about 469 arcmin$^2$ depth of these imaging data varies from about 28 to 30 magnitudes. The JADES fields are also covered by HST observations, and the JADES DR5 catalog provides HST photometry when available. We obtain the HST Adnanced Camera for Survey (ACS) images of objects from DAWN JWST Archive (DJA)\footnote{https://dawn-cph.github.io/dja/index.html}, which were reduced using \texttt{grizli} \cite{grizli, grizlizenodo}.

The spectra used in this work are obtained from the DJA, which provides publicly available JWST NIRSpec/MSA spectra uniformly reduced by the code \texttt{msaexp} \cite{msaexp}. For the spectrum of JADES-GS-217704, we multiply the uncertainty by a factor of 1.48. This factor is motivated by the \texttt{Prospector} SED fitting model. Specifically, we compute the difference between the best-fit SED model and the observed spectrum at $0.6{\mu{\rm m}}<\lambda_{\rm obs}<1.5{\mu{\rm m}}$, excluding the $0.8{\mu{\rm m}}<\lambda_{\rm obs}<1.2{\mu{\rm m}}$ window that is contaminated by the transmission signal and the Ly$\alpha$ damping wing. We then compute the noise scaling factor so that the reduced $\chi^2$ in the selected window equals one. This procedure is consistent with previous studies, which found that the error of NIRSpec MSA spectra is usually underestimated.

\begin{figure} 
	\centering
	\includegraphics[width=1\textwidth]{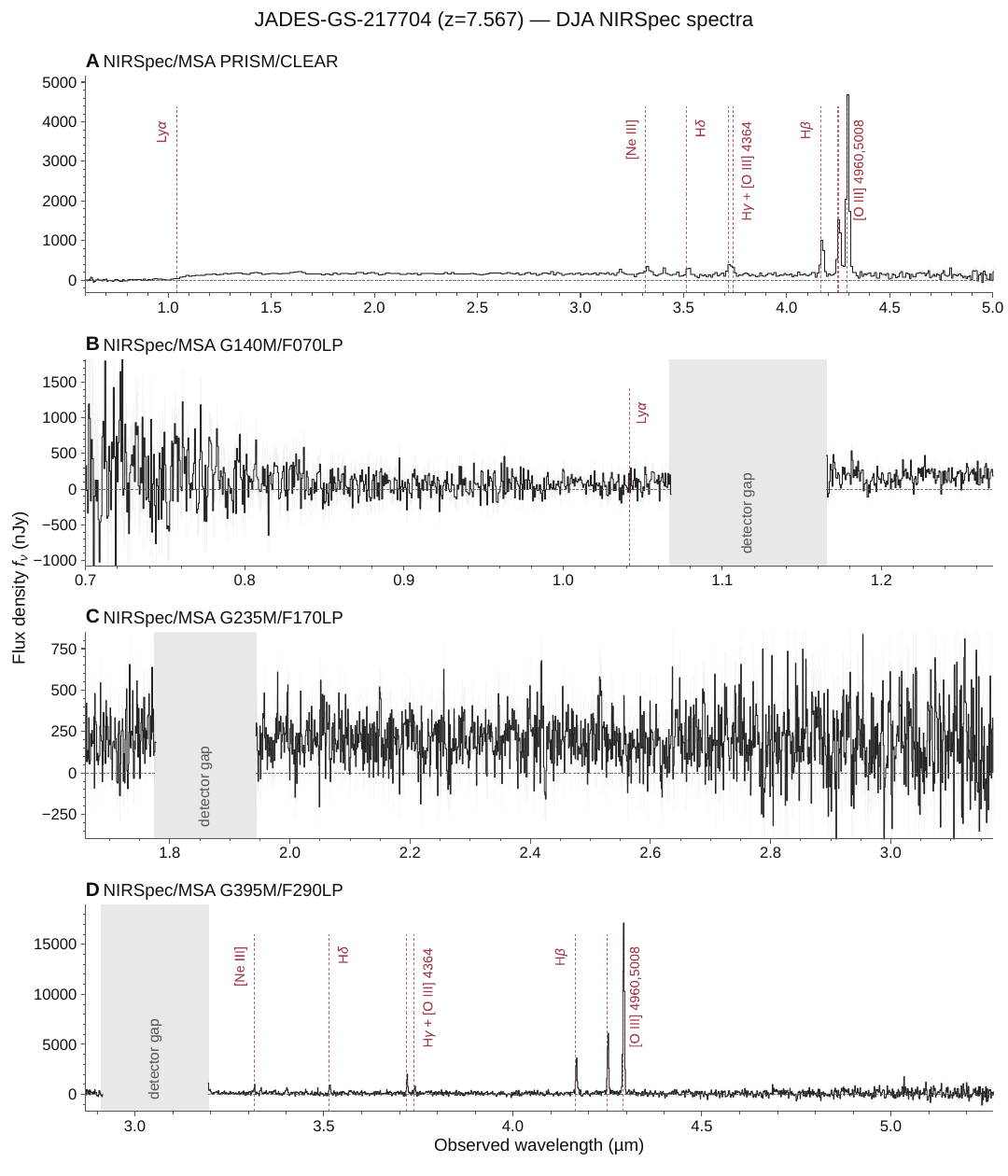} 
	\caption{{\bf The JWST NIRSpec/MSA spectroscopy for JADES-GS-217704.}  (\textbf{A}) the Prism spectrum. (\textbf{B}) the F070LP/G140M spectrum. (\textbf{C}) the F170LP/G235M spectrum. (\textbf{D}) the F290LP/G395M spectrum. The redshift of this galaxy is secruely determined by multiple emission lines.} \label{fig:allspec}
\end{figure}

JADES-GS-217704 is also observed by NIRSpec/MSA using the F070LP/G140M, F170LP/G235M, and F290LP/G395M setups. Figure \ref{fig:allspec} shows all the NIRSpec observations for JADES-GS-217704 available at DJA. The H$\beta$ and {[O\,\sc{iii}]} emission lines are clearly detected.

\subsubsection*{The FM-JADES-v1 Model and Population-Conditioned Outlier Search}

We use the self-supervised foundation model, FM-JADES-v1 \cite{ding26learningjwst1}, to perform the Population-Conditioned Outlier Search for high-redshift galaxies. FM-JADES-v1 takes JADES DR5 NIRCam images and photometric catalogs as inputs, tokenizes these data, and feeds these tokens into a Transformer block to learn the relations between the tokens. We refer the readers to \cite{ding26learningjwst1} for the architecture and training of the model. 

In this work, we use the frozen FM-JADES-v1 model to compute the CLS embedding vector for JADES objects. Following \cite{ding26learningjwst1}, we focus on a subsample of JADES objects, which have (1) coverage by F090W, F115W, F150W, F200W, F277W, F356W and F444W imaging; (2) signal-to-noise ratio (SNR) higher than 10 in the F356W and F444W bands. We only use image and catalog tokens relevant to these seven broad bands as inputs to produce the CLS embedding vector, which avoids the embedding vector encoding survey coverage or depth information rather than astrophysical information. The corresponding sample contains 122,336 objects. We then project the CLS embedding vectors onto a two-dimensional UMAP. The structure of this UMAP projection has been discussed in \cite{ding26learningjwst1}; this work focuses on its application in identifying high-redshift galaxies with unusual properties.

\begin{figure} 
	\centering
	\includegraphics[width=1\textwidth]{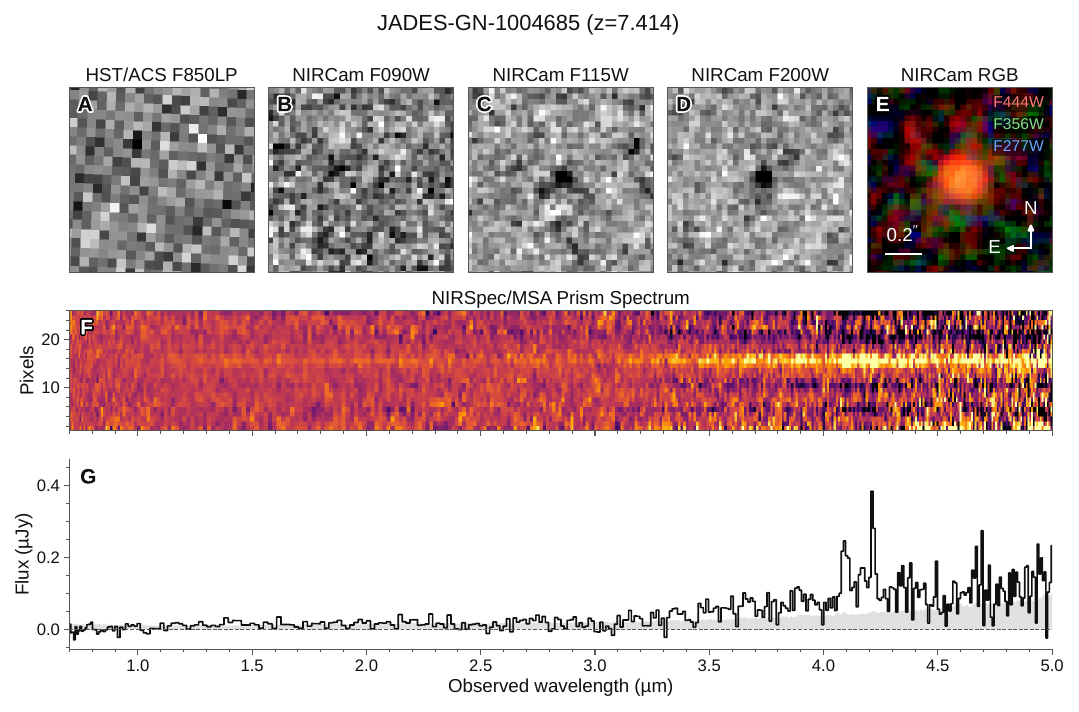} 
	\caption{{\bf HST and JWST observations for JADES-GN-1004685.} This object is a previously reported LRD (e.g., \cite{Rinaldi2025}), explaining the reason for it to be an outlier.  (\textbf{A}) the HST F850LP imaging. (\textbf{B}) NIRcam F090W imaging. Both of these bands show clear non-detections. (\textbf{C-D}) NIRcam F115W and NIRcam F200W images. (\textbf{E}) NIRcam F277W, F356W, and F444W color image. (\textbf{F} and \textbf{G})) NIRSpec/MSA Prism Spectrum.} \label{fig:outlier1}
\end{figure}

\begin{figure} 
	\centering
	\includegraphics[width=1\textwidth]{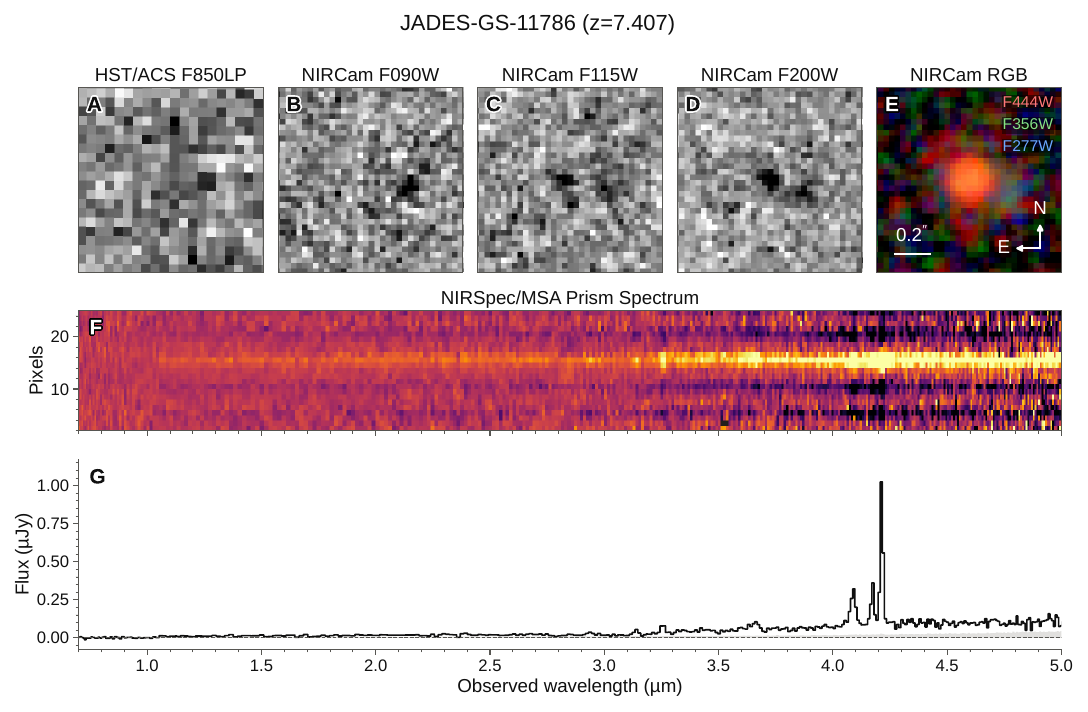} 
	\caption{Same as Figure \ref{fig:outlier1}, but for outlier JADES-GS-11786, which is a published LRD (e.g., \cite{Rinaldi2025}).}
	\label{fig:outlier2} 
\end{figure}

\begin{figure} 
	\centering
	\includegraphics[width=1\textwidth]{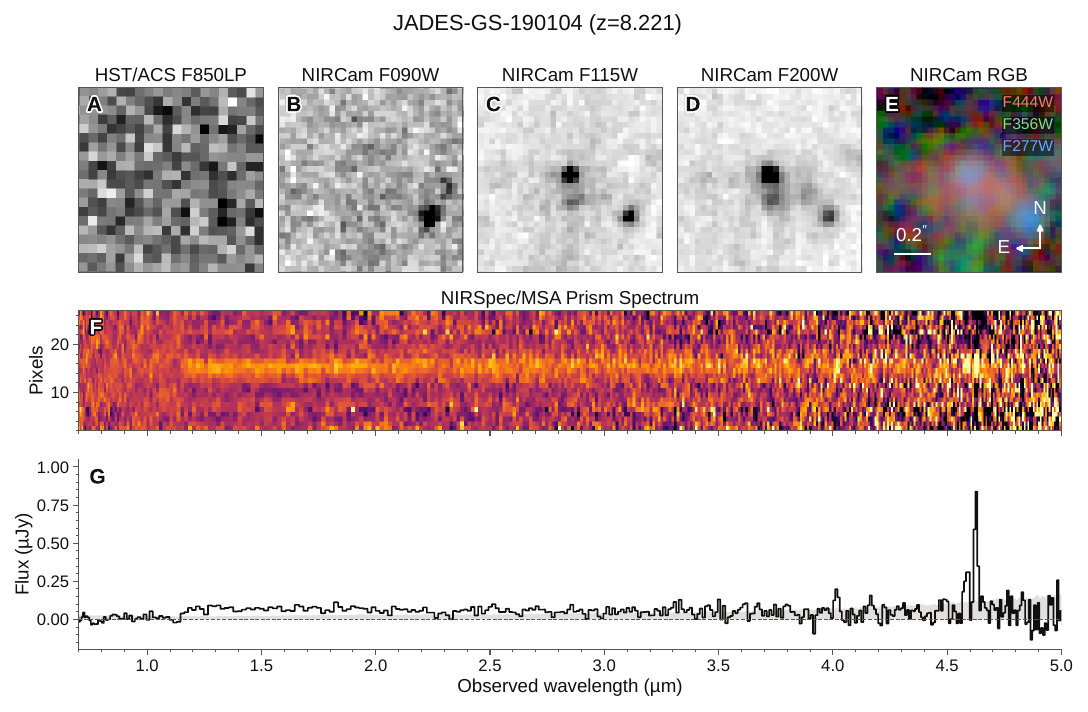}
	\caption{{Same as Figure \ref{fig:outlier1}, but for outlier JADES-GS-190104.} This galaxy has clear companions and nearby foreground contaminant.}
	\label{fig:outlier3} 
\end{figure}

\begin{figure} 
	\centering
	\includegraphics[width=1\textwidth]{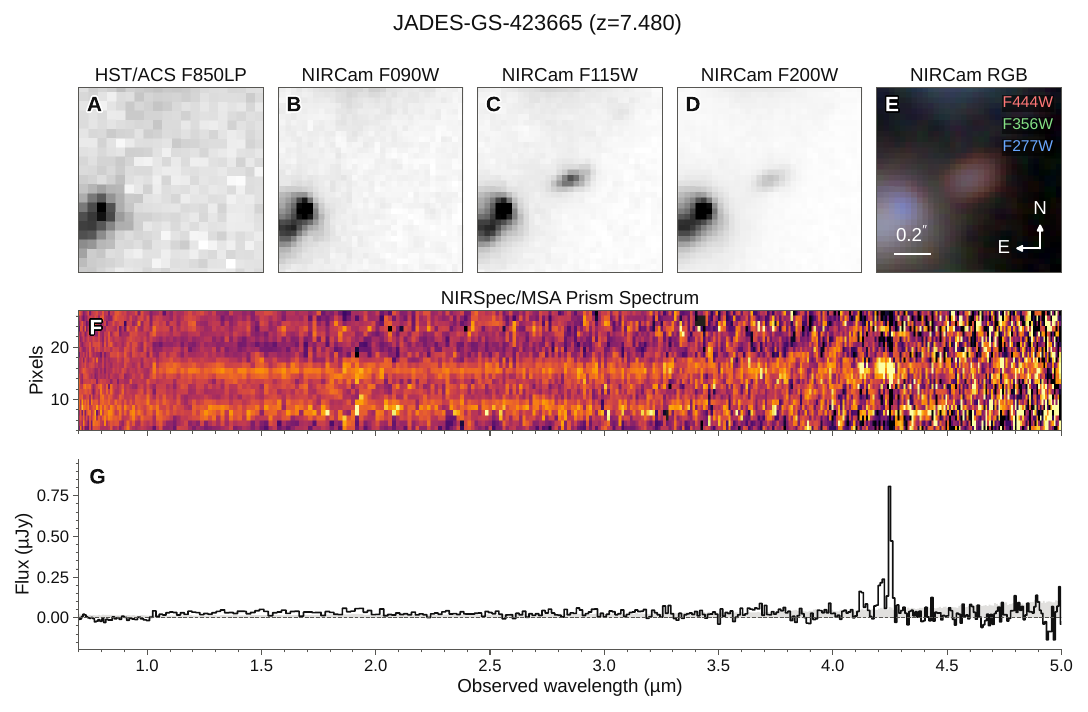} 
	\caption{Same as Figure \ref{fig:outlier1}, but for outlier JADES-GS-423665. This galaxy has bright foreground contaminants nearby.}
	\label{fig:outlier4} 
\end{figure}

To perform the Population-Conditioned Outlier Search, we start from $z>7$ galaxies in the JADES DR4 spectroscopic dataset \cite{CurtisLake2026,Scholtz2026}, where we require grade A or B for redshift determination. This criterion identifies 100 galaxies with $z>7$ among the 122,336 objects. We obtain the embedding vector of this sample, compute the median position of the 100 galaxies on the UMAP, then compute the distance from this median to each galaxy and rank the galaxies using these distances. The top five outliers are highlighted in Figure \ref{fig:method}. The main text discusses JADES-GS-217704 and the ionized superstructure FL$\alpha$RE-z7 on its sightline. Figure \ref{fig:outlier1} to Figure \ref{fig:outlier4} show the other four outliers identified by this framework.

\subsubsection*{The probability of a foreground interloper}

Before interpreting the transmission signal as an ionized structure, we carefully consider the possibility of a foreground interloper. In this scenario, JADES-GS-217704 does not have any flux bluer than its Ly$\alpha$ due to IGM attenuation, and a foreground object gives the signal we see at $\lambda_{\rm obs}\approx 0.95\mu$m. The F090W signal is entirely contributed by the foreground interloper.

We evaluate the number density of possible interlopers as follows. Using the {\texttt{CIRC3}} magnitudes in the JADES DR5 catalog, we find that JADES-GS-217704 has non-detections (below 2$\sigma$ levels) in the HST F435W, F606W, F775W, and F850LP bands. and a signal at $\approx3.5\sigma$ level in the F814W band. We note that F814W covers part of the positive signal at $\lambda_{\rm obs}>0.9\mu$m, and a marginal detection is plausible. Correspondingly, we select objects in the JADES catalog with $<2\sigma$ signals in F435W, F606W, F775W, and F850LP, as well as $<4\sigma$ signals in F814W. Since JADES-GS-217704 has {\texttt{CIRC3}} $m_{\rm F090W}=28.20\pm0.08$, we further require the possible interloepr to have $m_{\rm F090W}<28.3$. These criteria select $N=1,547$ objects in the entire JADES footprint. Assuming a maximum projected distance of $\Delta \theta=0.1''$ between the interloper and the background galaxy, the chance for a background galaxy to hit such an interloper is 

\begin{equation}
    P({\rm interloper})=N\pi\Delta\theta^2/A=2.87\times10^{-5}
\end{equation}

where $A=469{\rm ~arcmin}^2$ is the area of the JADES footprint.

Another argument against the interloper scenario is the spectral shape at $\lambda_{\rm obs}<1\mu$m. The positive flux peaks at $\lambda_{\rm obs}\approx0.95\mu$m and drops towards both shorter and longer wavelengths, reaching about zero at $\lambda_{\rm obs}\approx0.9\mu$m and $1\mu$m. If this feature is an emission line, the inferred line width is $2.3\times10^4{\rm ~km~s}^{-1}$ according to the Gaussian fit in Figure \ref{fig:size}, which is too large even for broad-line AGNs. If the positive flux is from galaxy continuum, then the spectrum at $\lambda_{\rm obs}<1\mu$m should be relatively flat, which cannot explain the gap at $\lambda_{\rm obs}\sim1\mu$m. 
We also search for foreground emission lines or absorption lines in the spectrum $\lambda_{\rm obs}>1\mu$m, and find no evidence for a foreground galaxy.

The spectral features, together with the low interloper probability, essentially rule out the foreground interloper scenario.

\subsubsection*{The wavelength-dependent morphology of JADES-GS-217704}

JADES-GS-217704 exhibits clumpy structures, and its morphology varies with wavelengths. One noticeable feature is the northern clump visible in the F115W and F150W images; this feature is not visible in the F090W image. In this section, we show that the wavelength-dependent morphology of JADES-GS-217704 can be explained by differential extinction.

 To this end, we perform pixel-by-pixel SED fitting for the galaxy using the PhotoIFU package \texttt{PhotoIFU} \cite{zhu2026photoifu}. In short, \texttt{PhotoIFU} takes NIRCam images as inputs, then performs SED fitting using \texttt{Prospector} for each pixel to get spatially resolved galaxy properties. All input NIRCam images are matched to the point spread function (PSF) of the F444W band; the PSF-matched images are obtained from JADES DR5. We exclude the F090W image from the fitting as standard \texttt{Prospector} runs cannot model IGM transmission like FL$\alpha$RE-z7. To model possible different extinction curves in different regions (e.g., \cite{nakazato_clump-scale_2026}), we make the following modifications to \texttt{PhotoIFU}. We use \texttt{dust\_type=4} in \texttt{Prospector}, corresponding to a power-law modification to the Calzetti dust attenuation curve \cite{kc13extinction}, and allow the extinction curve power law index to vary across pixels. We also include a \texttt{dust1} component that only attenuates emission from young stars (with $t_{\rm age}<10$ Myr).

\begin{figure} 
	\centering
	\includegraphics[width=1\textwidth]{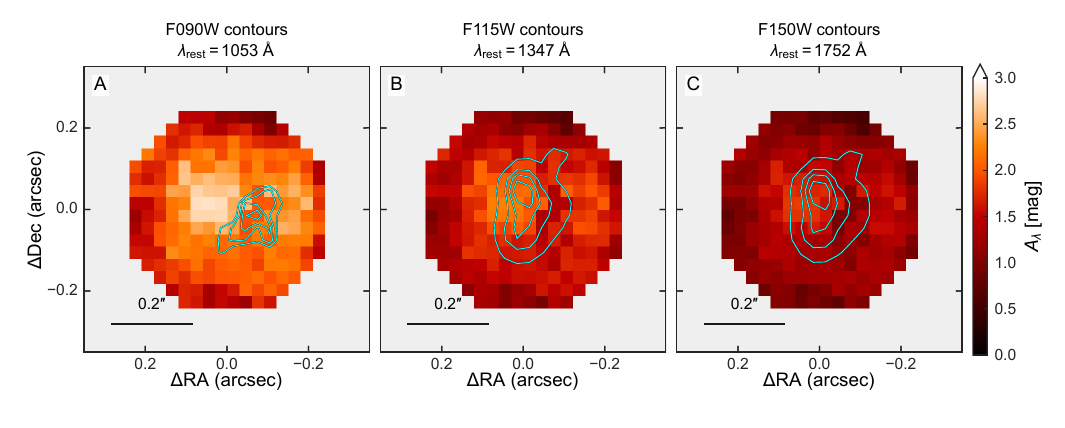} 
	\caption{The attenuation map of JADES-GS-217704 in three NIRCam short wavelength bands, derived from pixel-by-pixel SED fitting using \texttt{PhotoIFU} \cite{zhu2026photoifu}. The map shows a high extinction region at the northeast corner and a low extinction ``hole'' near the center of the galaxy. The contours show the surface brightness in each band. The pixel-by-pixel SED fitting clearly indicates differential reddening across the galaxy. The F090W has the strongest extinction, with the flux dominated by the low-extinction region.}
	\label{fig:dust} 
\end{figure}

Figure \ref{fig:dust} shows the overlay between the best-fit dust extinction map $(A_\lambda)$ and the surface brightness contours in the F090W, F115W, and F150W images. The F090W transmission is located exactly at the position where the dust attenuation is the lowest ($A_\lambda\lesssim1$ for the F090W filter). Meanwhile, the northern clump is located in the region with the largest attenuation, which has $A_\lambda\gtrsim2.5$ in the F090W band. We note that the northern clump contributes to $\approx20\%$ of the total flux in the F115W and the F150W images. 
A non-detection for the northern clump in the F090W band is consistent with its high dust attenuation and the integrated flux for the entire galaxy ($m_{\rm F090W}=28.2\pm0.08$, a 13$\sigma$ detection) in the F090W band.
Therefore, we suggest that JADES-GS-217704's wavelength-dependent morphology is a result of differential extinction \cite{nakazato_clump-scale_2026}.

\subsubsection*{Characterizing the transmission signal}

We measure the redshift and size of the transmission signal by fitting a Gaussian profile for the one-dimensional spectrum between 0.9$\mu$m and 1$\mu$m. We use the LSF from \cite{adg24} in the fiducial fit shown in Figure \ref{fig:size}. The model has three free parameters: the central wavelength $\lambda_{\rm c}$, the amplitude $A$, and the FWHM of the transmission signal. We use the nested sampling algorithm Nautilus \cite{nautilus} to obtain the posterior of these parameters, where we adopt flat priors for all parameters. The fit yields $A=0.0474^{+0.0094}_{-0.0095}\mu$Jy, $\lambda_{\rm c}=0.948\pm0.008\mu{\rm m}$, ${\rm FWHM}=0.07\pm0.02\ \mu\mathrm{m}$. The best-fit model is shown in Figure \ref{fig:size}.

\begin{figure}
    \centering
    \includegraphics[width=1\linewidth]{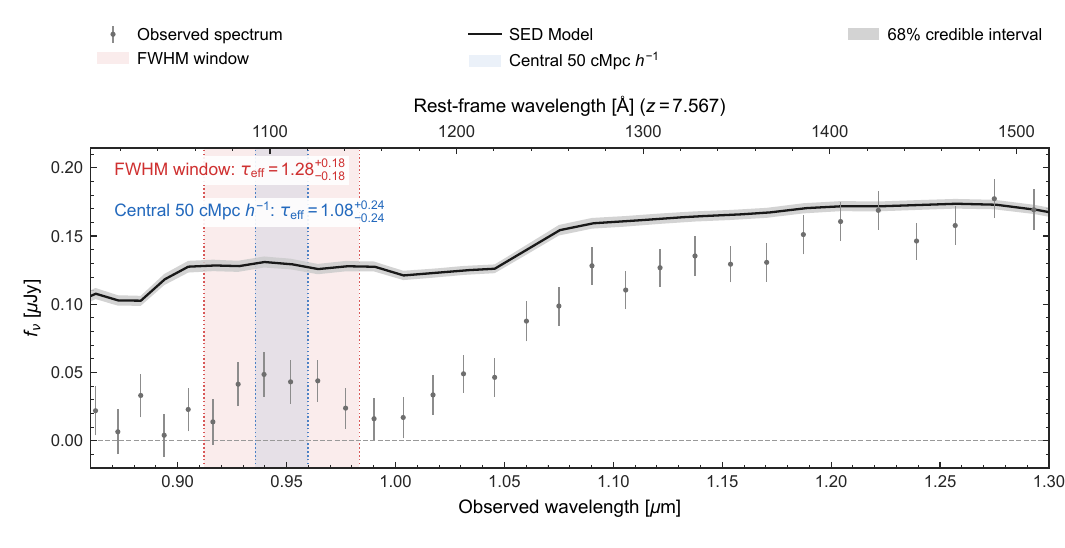}
    \caption{{\bf Continuum fitting and optical depth measurement for FL$\alpha$RE-z7.} The black solid line shows the best-fit SED model from {\texttt{Prospector}}, with IGM attenuation turned off. The gray shaded area marks the 1$\sigma$ error. Using this model, we evaluate the optical depth of FL$\alpha$RE-z7 in the FWHM-width window and the central 50 cMpc$/h$ window, yielding $\tau_{\rm eff}=1.28\pm0.18$ and $\tau_{\rm eff}=1.08\pm0.24$, respectively.}
    \label{fig:cont}
\end{figure}

We also evaluate the optical depth of FL$\alpha$RE-z7. To this end, we fit the galaxy's SED using \texttt{Prospector} \cite{prospector}. The input data includes the {\em HST} ACS, {\em JWST} NIRCam, and {\em JWST} MIRI \texttt{CIRC3} photometry provided by the JADES DR5 photometric catalog, as well as the NIRSpec/MSA Prism spectrum. We exclude the F090W magnitude and the Prism spectrum at $0.9\mu{\rm m}<\lambda_{\rm obs}<1.2\mu{\rm m}$ to avoid the influence of the transmission signal and the Ly$\alpha$ damping wing. Figure \ref{fig:cont} shows the best-fit SED and its posterior distribution. By turning off the IGM attenuation for the fitted SED, we can get the intrinsic continuum at $\lambda_{\rm obs}<1\mu$m.


Using the modeled intrinsic continuum, we measure the optical depth of FL$\alpha$RE-z7
using the wavelength range $0.912\mu{\rm m}<\lambda_{\rm obs}<0.983\mu{\rm m}$, corresponding to $|z-z_{\rm c}|<{\rm FWHM}/2$.
We compute the transmission flux and the modeled unattenuated continuum flux by summing the spectrum within the window. The effective optical depth is computed as 

\begin{equation}
    \tau_{\rm eff}=-\ln\langle T\rangle=-\ln(F_{\rm transmission}/F_{\rm cont}),
    \label{eq:tau}
\end{equation}

We compute the posterior distribution of $\tau_{\rm eff}$ using the posterior distributions of $F_{\rm cont}$ and $F_{\rm transmission}$, yielding $\tau_{\rm eff}=1.28\pm0.18$. We convert the optical depths to neutral fractions using the Gunn-Peterson relation (e.g., \cite{gunnpeterson,fan06}):

\begin{equation}
    \tau_{\rm eff}=1.8\times10^{5}h^{-1}\Omega_m^{-1/2}\left(\frac{\Omega_bh^2}{0.02}\right)\left(\frac{1+z}{7}\right)^{3/2}x_{\rm HI}
    \label{eq:tau}
\end{equation}

We use $\Omega_bh^2=0.0224$ from \cite{planck18}. This gives $x_{\rm HI}=(2.07^{+0.29}_{-0.30})\times10^{-6}$ for FL$\alpha$RE-z7.
We also perform the same measurements for the 50 cMpc$/h$ window around the central wavelength, yielding $\tau_{\rm eff}=1.08\pm0.24$ and $x_{\rm HI}=(1.74^{+0.38}_{-0.39})\times10^{-6}$.

We note that the transmission at $z\approx1\mu$m is consistent with zero, indicating that FL$\alpha$RE-z7 is not powered by the background luminous galaxy JADES-GS-217704 at $z=7.567$. The distance from JADES-GS-217704 to the edge of FL$\alpha$RE-z7 (i.e., from $z=7.567$ to $z=7$) is about 185 cMpc, much larger than the proximity zone sizes of the most luminous quasars \cite{Eilers2017}. The absence of the Ly$\alpha$ emission line and the existence of a Ly$\alpha$ damping wing for JADES-GS-217704 further reinforce this conclusion.

\subsubsection*{The tension between FL$\alpha$RE-z7 and cosmological simulations}

\begin{figure} 
	\centering
	\includegraphics[width=1\textwidth]{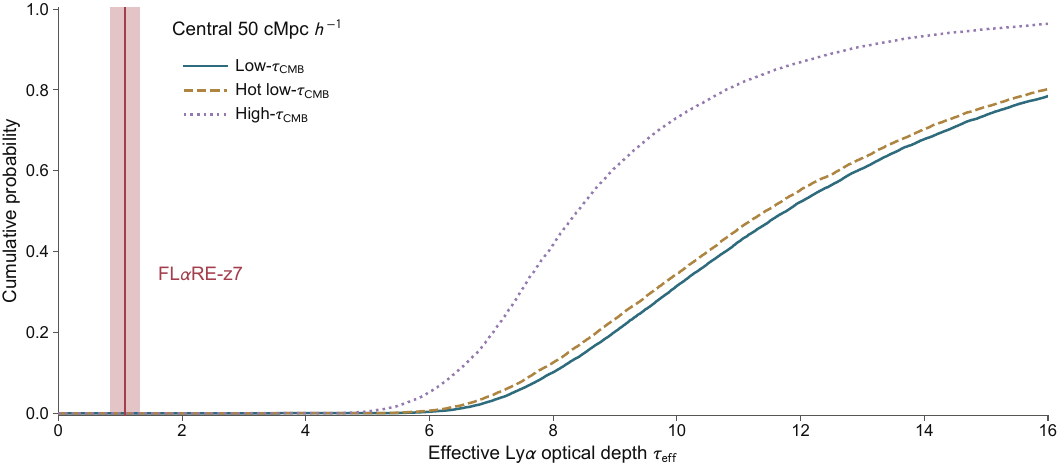} 

	\caption{\textbf{The cumulative distribution of $\tau_{\rm eff}$ from cosmological simulations.} The models are from \cite{keating20reionization}, corresponding to different reionization model parameters. The optical depths are measured for 50 cMpc$/h$ windows. None of the models produce a sightline with $\tau_{\rm eff}<4$, while FL$\alpha$SH-z7 gives $\tau_{\rm eff}=1.08\pm0.24$. Therefore, FL$\alpha$SH-z7 gives at least a marginal tension with current cosmological simulations and reionization models.}
	\label{fig:cdf} 
\end{figure}

We compare the evaluated $\tau_{\rm eff}$ of FL$\alpha$RE-z7 with \cite{keating20reionization}, who simulated the IGM optical depth for three reionization models with different ionizing photon energy and CMB optical depth: the low $\tau_{\rm CMB}$ model with $\tau_{\rm CMB}=0.051$ and $E_\gamma=17.1{\rm eV}$, the high $\tau_{\rm CMB}$ model with $\tau_{\rm CMB}=0.071$ and $E_\gamma=17.1{\rm eV}$, and the hot low $\tau_{\rm CMB}$ model with $\tau_{\rm CMB}=0.051$ and $E_\gamma=18.6{\rm eV}$. We refer to \cite{keating20reionization} for a detailed setup of these models. For each model, \cite{keating20reionization} provides 20,000 skewers, and we obtain the optical depths for the simulated skewers in the 50 cMpc$/h$ window centered at $z_{\rm c}$. None of these skewers have $\tau_{\rm eff}<4$. In other words, the simulations fail to produce sightlines like FL$\alpha$RE-z7 (Figure \ref{fig:cdf}).

We further evaluate the probability of finding a sightline like FL$\alpha$RE-z7 in the simulated skewers given the measurement errors. To do this, we obtain simulated spectra by applying the simulated IGM transmission to the best-fit SED model, then add Gaussian random noise to the simulated spectrum using the observed spectrum error. We then compute the transmission signal in the 50 cMpc$/h$ window centered on $z_{\rm c}$ for the simulated spectra. We choose the size of 50 cMpc$/h$ to match previous studies of reionization-era IGM (e.g., \cite{eilers_opacity_2018}). We estimate the chance that a random simulated sightline produces the observed spectrum of FL$\alpha$RE-z7 to be $9.14\times10^{-6}$, $9.16\times10^{-6}$, $5.00\times10^{-5}$ for the low $\tau_{\rm CMB}$ model, the hot low $\tau_{\rm CMB}$ model, and the high $\tau_{\rm CMB}$ model, respectively. Since FL$\alpha$RE-z7 is discovered from a sample of 100 spectroscopically confirmed $z>7$ galaxies, the probability of finding such a system in our entire experiment is $\approx5\times10^{-3}$. Therefore, FL$\alpha$RE-z7 is in marginal tension with the reionization models in \cite{keating20reionization}. The small probabilities evaluated using the three models indicate that this tension cannot be solved by early-reionization models or a higher ionizing photon energy.

Our result suggests that reionization might have larger spatial variations at $\gtrsim100$ cMpc scales compared to previous theoretical models. Recent observations of the ``dark gaps'' in high-redshift quasar spectra strengthen this argument, which are large neutral IGM structures at the end of EoR, when the Universe has been largely ionized \cite{becker_evidence_2015,Davies2018LongGap,zhu2022darkgap}.

\subsubsection*{The environment around FL$\alpha$RE-z7}

To investigate the environment around FL$\alpha$RE-z7, especially whether it resides in a galaxy overdensity, we evaluate object number densities in the GOODS-S field and around the FL$\alpha$RE-z7 sightline. To prevent the heterogeneous coverage of JWST observations in the JADES fields from biasing the measurements, we construct a depth-limited footprint as follows. We first compute the $5\sigma$ depth maps of GOODS-S by converting the ERR extensions of the mosaic using the following equation:

\begin{equation}
	m^{\rm lim}=31.4 - 2.5 \log(5\sigma_{\rm aper})=31.4 - 2.5 \log\left(5\sqrt{N_{\rm pix}}\times{\rm ERR}\right)
	\label{eq:depth} 
\end{equation}

where 31.4 is the magnitude zeropoint when using nJy as flux units, and $N_{\rm pix}$ is the aperture for magnitude measurement. Here we use apertures with $r=0.1''$, yielding $N_{\rm pix}=\pi(r/0.03'')^2=34.9$. We neglect the noise correlation between pixels, as our primary goal is to select a depth-limited coverage footprint, rather than accurately compute the depth map. We then construct the depth-limited footprint by selecting pixels with the following magnitude limits: $m_{\rm F090W}^{\rm lim}>28.887$, $m_{\rm F115W}^{\rm lim}>28.902$, $m_{\rm F150W}^{\rm lim}>28.790$, $m_{\rm F200W}^{\rm lim}>29.011$, $m_{\rm F277W}^{\rm lim}>30.169$, $m_{\rm F356W}^{\rm lim}>30.200$, $m_{\rm F444W}^{\rm lim}>29.941$. These depths are determined as the 1st percentile of all pixels that have $d<2'$ away from the FL$\alpha$RE-z7 sightline. This results in a footprint of 149 arcmin$^2$ in GOODS-S, which is shown as the gray area in Figure \ref{fig:environ}.

We then select objects with $|z_{\rm phot}-z_{\rm c}|<0.2$ in the GOODS-S field, using the JADES DR5 photometric catalog. We only select objects that are brighter than the magnitude limits listed above, except for the F090W magnitudes, which is a dropout band for $z\gtrsim7$ galaxies. This forms a flux-limited sample on the corresponding depth-limited footprint. We find 214 objects across the 149 arcmin$^2$ footprint, yielding an object number density of $1.44\pm0.10{\rm ~arcmin^{-2}}$ (assuming Poisson noise). There are 27 objects within 2' around FL$\alpha$RE-z7, corresponding to $2.28\pm0.44{\rm ~arcmin^{-2}}$. In other words, we do not find significant overdensities around FL$\alpha$RE-z7.
We perform the measurement using different depths and aperture sizes around FL$\alpha$RE-z7, and the conclusion remains unchanged.

We note several limitations for our analysis. First, we use photometric redshift in our analysis, as only a small subset of galaxies around FL$\alpha$RE-z7's sightline have spectroscopic redshifts. Second, the object sample we use to compute surface number densities is flux-limited, and the abundance of faint galaxies $(M_{\rm UV}\lesssim-17)$ around FL$\alpha$RE-z7 is unknown. Those faint galaxies might be the primary contributor to reionization (e.g., \cite{atek24reionization}), and we cannot rule out the scenario that FL$\alpha$RE-z7 resides in an overdensity of faint galaxies.

\subsubsection*{Aritficial Intelligence Disclosure}

We use AI to aid the coding in this study, including GPT-6 Astra \cite{openai_chatgpt_2026}, Claude Opus 5, Opus 5.5, Fable 5, and Fable 5.1 \cite{anthropic_claude_2026}. We also use these AI to improve the writing of the manuscript.






\newpage






\bmhead{Acknowledgements}



We acknowledge the independent discovery of JADES-GS-217704 by J. Helton et al. \cite{helton2026LAF}.
We sincerely appreciate the open communication and helpful discussion from the JADES team.

This work is based on observations made with the NASA/ESA/CSA James Webb Space Telescope. The data were obtained from the Mikulski Archive for Space Telescopes at the Space Telescope Science Institute, which is operated by the Association of Universities for Research in Astronomy, Inc., under NASA contract NAS 5-03127 for JWST. 

Some of the data products presented herein were retrieved from the Dawn JWST Archive (DJA). DJA is an initiative of the Cosmic Dawn Center (DAWN), which is funded by the Danish National Research Foundation under grant DNRF140.

The authors used generative AI to improve the writing of the manuscript. Specific models used include GPT-6 Astra \cite{openai_chatgpt_2026}, Claude Opus 5, Opus 5.5, Fable 5, and Fable 5.1 \cite{anthropic_claude_2026}.

\begin{itemize}
\item Funding

M.Y.\ acknowledges support from the Steward Observatory, the University of Arizona through the Bart J. Bok Fellowship.
Y.Z.\ acknowledges support from the NIRCam Science Team contract to the University of Arizona, NAS5-02105. Y.Z. is also supported by JWST Program \#6434. 

\item Conflict of interest/Competing interests

The authors declare no competing interests.

\item Ethics approval and consent to participate

\item Consent for publication

\item Data availability

The JADES DR5 dataset is available online at \url{https://slate.ucsc.edu/~brant/jades-dr5/}. The DJA is accessible at \url{https://dawn-cph.github.io/dja/}. Other data used in this work will be available online upon the acceptance of this paper.

\item Materials availability

\item Code availability

The code used in this work will be available online upon the acceptance of this paper.

\item Author contribution

MY applied the deep learning model for the discovery of the object, performed measurements for the object, and came up with the astrophysical interpretation.
YZ provided interpretations and modeling related to cosmic reionization in this work.
JD provided the deep learning model FM-JADES-v1 and provided suggestions in the application of this model in astronomical discovery.
All other authors provided suggestions in data analysis and manuscript preparation.

\end{itemize}

\noindent







\bibliography{sn-bibliography}

\end{document}